\documentclass{article}

\usepackage{PRIMEarxiv}
\usepackage{comment}

\usepackage[utf8]{inputenc} % allow utf-8 input
\usepackage[T1]{fontenc}    % use 8-bit T1 fonts
\usepackage[colorlinks=true, linkcolor=black, citecolor=black, urlcolor=black]{hyperref}
\usepackage{url}            % simple URL typesetting
\usepackage{booktabs}       % professional-quality tables
\usepackage{amsfonts}       % blackboard math symbols
\usepackage{nicefrac}       % compact symbols for 1/2, etc.
\usepackage{microtype}      % microtypography
\usepackage{lipsum}
\usepackage{fancyhdr}       % header
\usepackage{graphicx}       % graphics
\graphicspath{{media/}}     % organize your images and other figures under media/ folder

\title{

Smart UX-design for Rescue Operations Wearable - A Knowledge Graph Informed Visualization Approach for Information Retrieval in Emergency Situations
 \thanks{\textit{2023 IEEE International Conference on Electro Information Technology (eIT) | 978-1-6654-9376-5/23/\$31.00 ©2023 IEEE | DOI: 10.1109/EIT57321.2023.10187320
  }}}

\author{
  Mubaris Nadeem, Johannes Zenkert, Christian Weber, Madjid Fathi \\
  Institute for Knowledge-Based Systems and Knowledge Management \\
  University of Siegen\\
  Hoelderlinstrasse 3, 57068 Siegen, Germany\\
  \texttt{mubaris.nadeem@uni-siegen.de} \\
   \And
  Muhammad Hamza \\
  mbeder GmbH\\
  57068 Siegen, Germany
}

\begin{document}
\maketitle

\begin{abstract}
This paper presents a knowledge graph-informed smart UX-design approach for supporting information retrieval for a wearable, providing treatment recommendations during emergency situations to health professionals. This paper describes requirements that are unique to knowledge graph-based solutions, as well as the direct requirements of health professionals. The resulting implementation is provided for the project, which main goal is to improve first-aid rescue operations by supporting  artificial intelligence in situation detection and knowledge graph representation via a contextual-based recommendation for treatment assistance.
\end{abstract}

% keywords can be removed

\section{Introduction}

\subsection{The KIRETT Project}
A rescue operation is a rapid and risky emergency, in which health professionals must establish a fast and accurate medical supply for the person in need. In this situation, the time of action is limited and acting professionals are in constant need for up-to-date information, retrieved from available medical devices. Age, weight, blood pressure, pulse value, and ECG data are some of the crucial data, which are derived and need constant evaluation during medical treatment as they vary the treatment given to the patient. The KIRETT project provides a wearable device combining medical data processing with modern artificial intelligent tools to identify situations from real-time recorded data and subsequently provides a step-wise treatment recommendation, based on "treatment paths and standard operating procedures for the rescue service in the district of Siegen-Wittgenstein" \cite{b1}. This added recommendation, situation-aware and treatment-centered, goes beyond existing patient observation monitoring and actively supports health professionals in emergency situations. Warnings and alarms report drastic changes in monitored data and provide modern assistance for treatment and emergency management in a constantly changing environment. 

\subsection{Requirement Analysis} \label{requirments}
Vital for such a solution is the design and implementation of an efficient, visual component for the user, which is explicitly constructed for a medical emergency situation in which time, place, and environmental measurements, as well as information transfer limitations, are taken into consideration. 
As an efficient, situation-aware information transmission is crucial for effective recommendations, the KIRETT treatment knowledge graph is used to inform the design decisions, considering transitions in the graph as interface inputs, signs for prompts, options, and screen transitions. As such, the requirements for a health professionals-focused user interface were identified in regard to the treatment knowledge graph \cite{b1} and its inherent treatment support, as well as based on direct exchanges with the rescue personnel: 

\begin{itemize}
\item \textbf{Big visual interaction components, like buttons, arrows, icons}, are needed to navigate easily through the knowledge graph and to empower Yes/No question-nodes covering.  
\item \textbf{Warnings/Notification-screens} are needed to visually highlight individual graph-based recommendations or data changes. This also includes time-critical notifications for the end user.
\item \textbf{Approval screens} are needed to approve data-based decisions for selection branches in the treatment knowledge graph. It is also needed in case of different outcomes in situation detection, based on more data during an emergency case. 
\item \textbf{Navigational, graph-based overview screen} to consider treatment steps in the given emergency scenario. 
\item \textbf{Patient monitor} is needed to visualize patient data at hand.
\item \textbf{Massive information visualization} needed to visualize all treatment paths and additional information from the knowledge graph.
\item \textbf{Element-selecting screen} is needed to allow users to select situations from situation detection and treatment selection to decide on further wearable assistance. 
\end{itemize}

\section{Theoretical Background}
This section reviews the essential theoretical background information needed for this paper. 

\subsection{UX-Design}
User experience is described as a feeling or an effect, which occurs through previous, active, and post-interactions with an application. The UX-designer focuses on the goal to derive the end users' subjective feeling of satisfaction and fulfillment in the product at hand \cite{hartson2018ux}. Hartson and Pyla describe that there are four characteristics of user experience, which are the type of interaction, the totality of effects, the subjective feeling by the end user, and the context, in which the user is exposed to the product \cite{hartson2018ux}. The type of interaction is presented as a direct and indirect interaction, which differs from the active hands-on experience of the user, or the passive effects felt by the user. The totality of effects includes the point of usability and the unfolding of effects over time. The third characteristic describes the subjective feeling of the end user which may differ from person to person. Hartson and Pyla describe that the user context is essential for user experience so that the usage in the different environment is felt completely distinct from the user \cite{hartson2018ux}. In respect of the wearable designed and developed for rescue operations in the KIRETT project, the four characteristics will be described briefly: 
\begin{itemize}
    \item \textbf{Interaction:} 
    The wearable uses a direct interaction method with the user to assist in health rescue operations while using the knowledge graph to navigate through the treatment recommendations and get feedback for medical situations derived from connected medical data.  
    \item \textbf{Totality of effects:}
    The totality of effects is partially achieved by fostering usability during rescue operations to provide meaningful situation detection and recommendation in action. The unfolding over time is not used here, thus the  demonstrator developed unpacks its full potential from the beginning. 
    \item \textbf{Subjective user feeling:}
    The subjective user feeling is mainly the driver for the design and development of this demonstrator. Health professionals, acting in emergency situations, know what they need and expect from the system. Users from different fields may not understand the need for such a demonstrator, because of the very hands-on tailoring to the rescue personnel. 
    \item \textbf{Context:}
    The context was taken into consideration to develop a hands-on wearable for users who are working in the medical field during a rescue operation. 
\end{itemize}

\subsection{Knowledge Graph}
A knowledge graph is a representation of data that allows one to gather and process data. It is represented in nodes and relations, which describe the entity as a representation of knowledge and the relations which allow differentiation between entities\cite{b2}. Knowledge graphs are widely used in the fields of healthcare and medicine, e.g. electronic medical record \cite{rotmensch2017learning}, precision medicine \cite{chandak2023building}, and safe medicine recommendation \cite{gong2021smr}. In the KIRETT project, the developed treatment paths of the given actions in rescue operations are modeled and stored in a Neo4j\footnote{https://neo4j.com/} graph database and kept in the memory of the wearable device \cite{zenkert2022kirett}, where they can be accessed through cypher query (data retrieval language) requests on the graph database \cite{fernandes2018graph}.

The graph contains a variety of node types, e.g., jump and decision node, and contains thousands of relations and properties to cover all potential treatments and recommendations.
%, along the medical literature.  
The interactivity between the code base and the graph will be archived by the usage of Python, with a query-based concept \cite{zenkert2022kirett}, in which queries will be formed in real-time to provide the health professional with the needed information.

\subsection{Board and LCD Selection}

The project requires a device that has an FPGA component to be able to synthesize a tensor processor in its Programmable Logic (PL) for real-time utilization of an integrated neural network. Furthermore, the device has to be powerful enough to run an embedded \textit{Linux} operating system. In addition, the device has to support external peripherals such as an LCD touchscreen, Bluetooth, and Wi-Fi dongles.

Following that, the \textit{TE0802} by \textit{Trenz Electronic} has been selected. The \textit{TE0802} is a highly advanced FPGA module designed to offer high processing power and flexibility. It features a \textit{Xilinx Zynq UltraScale+ MPSoC}, with a dual-core \textit{ARM Cortex-A53} application processor, a dual-core \textit{ARM Cortex-R5} real-time processor, and an FPGA fabric capable of supporting up to 103,320 logic cells\cite{specs} \cite{xilinx_2023}. 

The \textit{PMOD MTDS} touch screen has been selected as the user interface. The module features a 2.8-inch capacitive touchscreen display with a resolution of 320 x 240 pixels and supports up to five simultaneous touch points for a responsive user interaction \cite{pmod}. The \textit{PMOD MTDS} is also equipped with a serial peripheral interface (SPI) interface for easy communication with a host system \cite{pmod}.

\subsection{Limitations of the LCD} \label{limitation}

While the \textit{PMOD MTDS} touchscreen by \textit{Digilent} offers a range of advanced features and capabilities, there are limitations to consider. The screen has a relatively small size of the display. Furthermore, there is only one font of size 10px available in its library which can not be changed or modified. The screen has a limited resolution of 240p. Another limitation is the module's low refresh rate, which limits applications that require high-speed graphics rendering. The screen has a low reaction time and  needs to replace the entire screen in order to display new content. Due to its low reaction time, there can be a noticeable delay between the user's touch input and the resulting action on the screen.

\section{Methodology}
To overcome the above-listed limitations, concepts, and methods were developed to surmount the challenges.

\subsection{INTER-PROCESSOR MEMORY MAPPED COMMUNICATION} \label{memorylocation}

The idea was to have a single processor manage all the processes and tasks of a computer system, including those responsible for displaying information. However, as the number of processes increases, the system may struggle to meet completion deadlines \cite{silberschatz_galvin_gagne_2011}, resulting in a glitching display. To address this and ensure a smooth and uninterrupted operation, a separate processor on the silicon will be assigned the task of handling display-related processes. 

As previously mentioned, the \textit{TE0802} board features two distinct processors - the \textit{ARM Cortex A53} and the \textit{ARM Cortex R5} \cite{xilinx_2023}. Initially, the \textit{A53} processor was responsible for all tasks on the board. However, after considering the advantages of utilizing the additional processor, it has been determined that tasks related to displaying will be executed on the \textit{ARM R5} processor. 
The communication between tasks will be executed by the \textit{A53} and \textit{R5} processors jointly. To implement this, the method of memory mapping has been utilized.

The display component uses two types of variables - static and dynamic. Static variables consist of unchanging information on the screen, such as buttons on a home screen or a welcoming title displayed when the LCD boots up. Dynamic variables can be updated over time, as in the case of displaying the heart rate value of a patient continuously.

\begin{figure}[htbp]
\centerline{\includegraphics[width=0.35\textwidth]{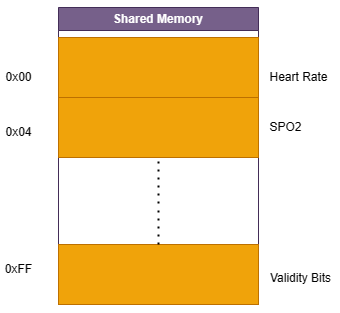}}
\caption{Shared Memory Partition}
\label{memory_mapping}
\end{figure}

The memory employed to facilitate communication between the processors is designed to store solely dynamic variables. Meanwhile, static variables can be initialized in the local volatile memory of the \textit{R5} processor. To map the variables to the memory, only primitive types of variables have been taken into account in this architecture. Once the variables have been defined, they are mapped to their specific memory location. In the case of dynamic variables, validity bits are defined (refer to figure \ref{memory_mapping}). Validity bits are toggled to valid upon entry of new data into the memory. Furthermore, they are set to invalid upon completion of the task.

The following example illustrates the visualization of \texttt{SpO2} on the LCD-Screen
as a pipeline:

\begin{enumerate}

\item A process on the \textit{A53 processor} reads the value of \texttt{SpO2} from Bluetooth-connected device.
\item The value is written as an integer value to its designated memory location, \texttt{0x04}, using Memory Mapped Input Output (MMIO).
\item The process also sets the validity bit of the \texttt{SpO2} variable in the memory.
\item An integer pointer in the \textit{R5} processor reads the value, referencing memory location \texttt{0x04}.
\item The integer value is then displayed on the screen, followed by unsetting the validity bit.
\end{enumerate}

\subsection{DYNAMIC LENGTH-BASED TEXT-TO-IMAGE-GENERATOR FOR MEMORY SAVING}
\begin{figure}[htbp]
\centerline{\includegraphics[width=0.35\textwidth]{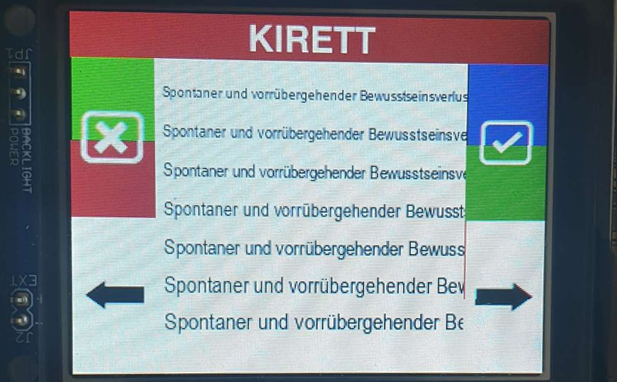}}
\caption{font size-test}
\label{font sizes}
\end{figure}

As described in the previous sections (see section \ref{limitation}), the used LCD is not capable of enhancing the font size, which results in minimal readable text. During a stressful emergency situation, the text and elements should be easy to read and handle. To overcome this situation, a bigger font size was needed to enhance the readability of the text. To accomplish this, a font size-test was established on the screen (fig. \ref{font sizes}) to test which font sizes can be used to allow sufficient readability.  

Due to different text lengths, a dynamic approach was needed to generate the text images. So a script was created to use different font sizes, matching the required text length with optimized readability.  In total 6653 images were generated in different colors, for the different navigational steps, and 6450 images were generated for the different treatment paths for rescue operations (German: "Behandlungspfade (BPR")), treatment groups, treatment categories (38 images), situation detection (10 images), alphabet with German-specific characters (58 images) and numbers in different colors (30 images). In addition to that, all buttons were designed as images. That resulted in 22 images for navigational purposes.

\begin{figure*}
\centerline{\includegraphics[width=0.8\textwidth]{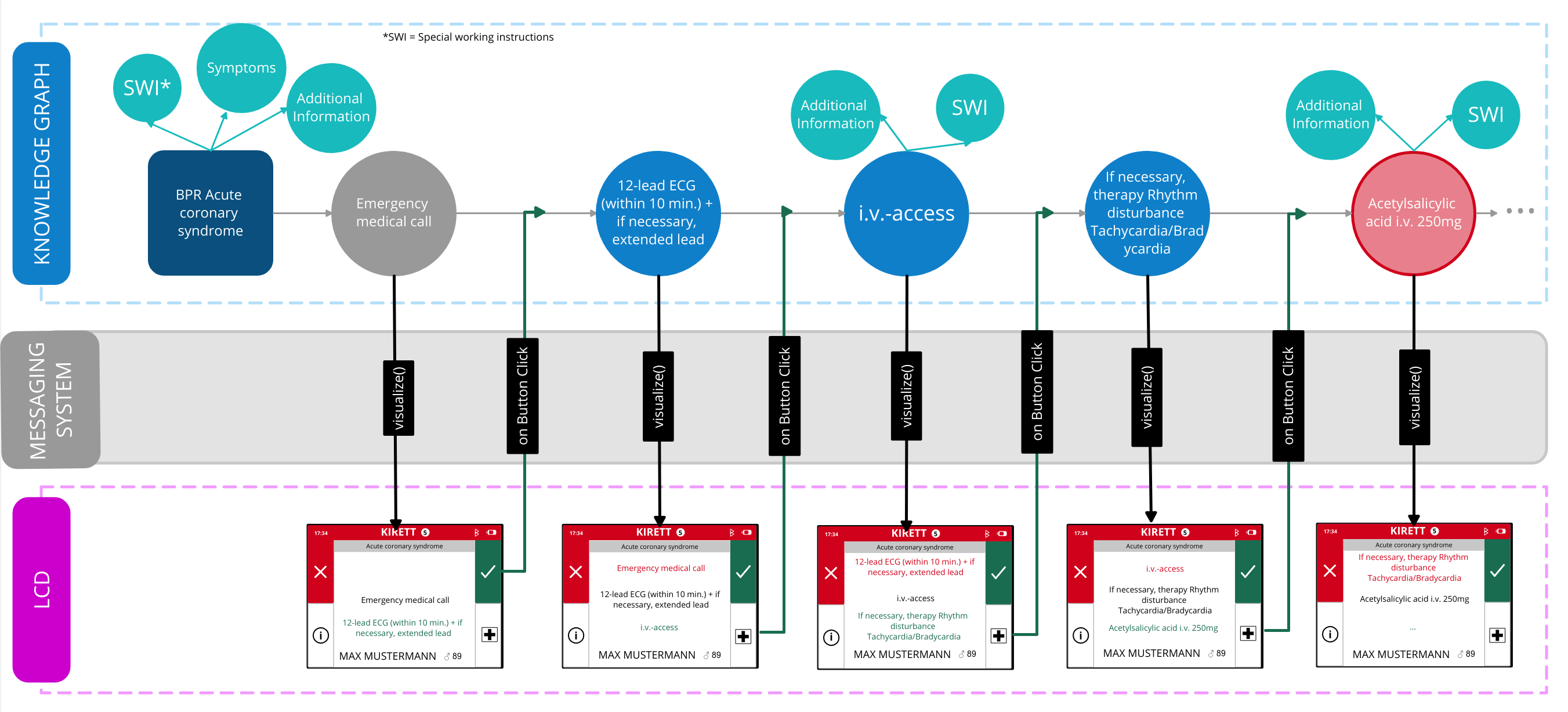}}
\caption{Graph and UI Interconnection }
\label{scheme}
\end{figure*}

\subsection{ILLNESS GROUP CONVENTION ID} \label{sd}

As addressed in Section \ref{memorylocation} all information that needs to be visualized, needs to be saved in the memory of the \textit{ARM R5} processor, to be directly accessible for the LCD component. Due to its minimal space, long text elements are not usable. For this, illness, accident, and disease groups were designed and mapped as short codes to the situation detection outcomes as \texttt{convention-id}. The ID was used to communicate the right \texttt{situation-detection-group} to the LCD and to the knowledge graph. 

\begin{table} [ht] \label{table2}
\caption{Convention IDs for situation detection and memory-communication}
\begin{center}
\begin{tabular}{|c | c |} 
 \hline
situation-detection-group & convention-id\\ [0.5ex] 
 \hline\hline 
ZNS-Krankheiten  & sdz\\ 
Herz-Kreislauf-Erkr.  & sdh\\ 
Erkr. der Atemwege  & sda\\ 
Erkr. des Bauchraums & sdb\\ 
Psychiatrische Erk.  & sdp\\ 
Stoffwechselkrankheiten & sds\\ 
Gyn.-geburtshilf. Notfaelle & sdg\\ 
Andere Krankheiten & sdak\\ 
Infektionen & sdi\\ 
Reanimation & sdr\\ 
\hline
\end{tabular}
\end{center}
\end{table}

\section{Graph integrated interface design}
In the following, the intermediate results of the UX-design are presented. \\

Figure \ref{scheme} presents an overview of the communication and design of the interconnection between graph and UX components. Communication is enabled with a messaging system.
The Neo4j knowledge graph is designed as a sequential graph, in which patient observation inputs from the interface and middleware-recorded medical data decide which next node is selected. Depending on the current node, more information can be visualized through buttons. For example patient data (fig. \ref{screens} (B)) with the patient monitor button (fig. \ref{scheme}, bottom-right button) or additional information, with the "provided" button (fig. \ref{scheme}, bottom-left button). The ongoing situation detection uses an artificial neural network to further integrate a vector of probabilities as a situation detection. The situation detection at hand is an artificial neural network that is used to detect various health situations, like cardiovascular or respiratory disease. For the evaluated complications, various algorithms were tested, and the selected ANN was trained with more than 300,000 previously recorded data from such rescue operations. During the rescue operation, the trained network is fed with freshly recorded data. It predicts a probability of complications with the data at hand, which is then presented to health professionals to choose the right complication group \cite{zenkert2022kirett}. The algorithm used was based on a previous study in which respiratory emergency situations were detected with the help of machine learning algorithms \cite{9968356}. The partners have tested multiple algorithms to find the best suitable machine-learning algorithm for each complication.

\begin{figure}[htbp]
\centerline{\includegraphics[width=0.40\textwidth]{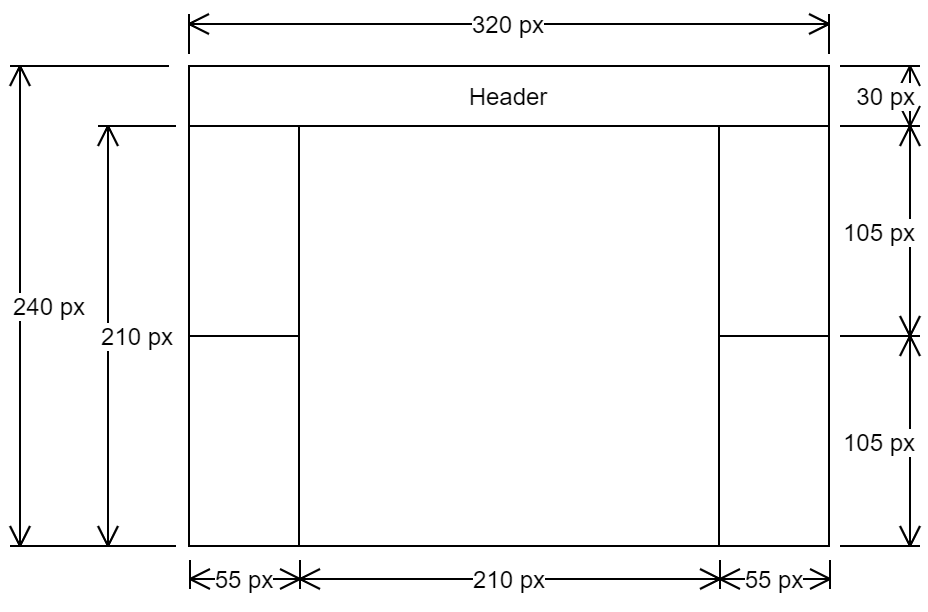}}
\caption{UI static pattern is presented in a grid layout with 6 different panels.}
\label{dimensions}
\end{figure}

A grid pattern with six different panels (see fig. \ref{dimensions}) was designed to have a static layout for the dynamic environment of the end users. The base panel integrates a header, which focuses on displaying the time, the main menu button (fig. \ref{screens} (A)), and the battery status. The sides are occupied by a total of four buttons, which can have different dynamic functions. The main information for medical data, situation detection, and treatment recommendations is visible in the central space between the four buttons. 

\subsection{Icon-based Visual Interaction Components}%, like buttons, arrows,  and icons} 
\label{buttons}
Due to protective gear of health professional in emergency situations, as e.g., gloves, each button, and visual interaction component needs to be large in size.
For that, a fixed size of button size of 105px x 55px (see fig. \ref{dimensions}) was used for the main four buttons. Buttons for the main menu were modified to fit to the minimum width of 55px. For main menu buttons a width of 64px was selected. Other visual components, like icons for a heartbeat, connected devices via Bluetooth, and a gender symbol were set with a minimal width of 23px.

\subsection{Warnings/Notification-Screen}
\begin{figure}[htbp]
\centerline{\includegraphics[width=0.30\textwidth]{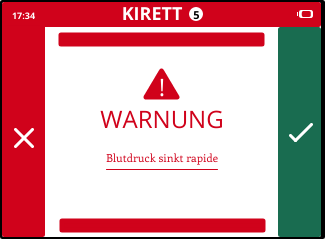}}
\caption{Warning and notification screen}
\label{warning}
\end{figure}

A warning/notification screen (fig. \ref{warning}) is used to visualize essential information about the treatment. This includes drastically changing vitals or time-related treatment-components, which needs to be done in a specific time period. The presented screen shows a symbol (warning or a notification), including more detailed information.

\subsection{Approval Screens}
Medical data must be actively fed into the knowledge graph environment, to enable decisions based on medically necessary data. For example, medical drug intake may differ for children of a certain age compared to adults \cite{b1}. In such a case the prompt/approval message screen is triggered and shows the current treatment state and the utilized information.
 
\begin{figure}[htbp]
\centerline{\includegraphics[width=0.30\textwidth]{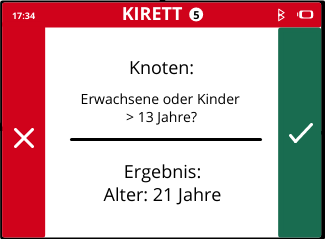}}
\caption{Dynamic prompt message-screen}
\label{promptmessage}
\end{figure}

Fig. \ref{promptmessage} presents a screen, which shows the current treatment node in the knowledge graph and the integrated requirement to have knowledge about the age  of the patient, e.g., "Adult or Child $>$ 13 Year?". 

\subsection{Navigational, graph-based Overview Screen}

\begin{figure*}[htbp]
\centerline{\includegraphics[width=\textwidth]{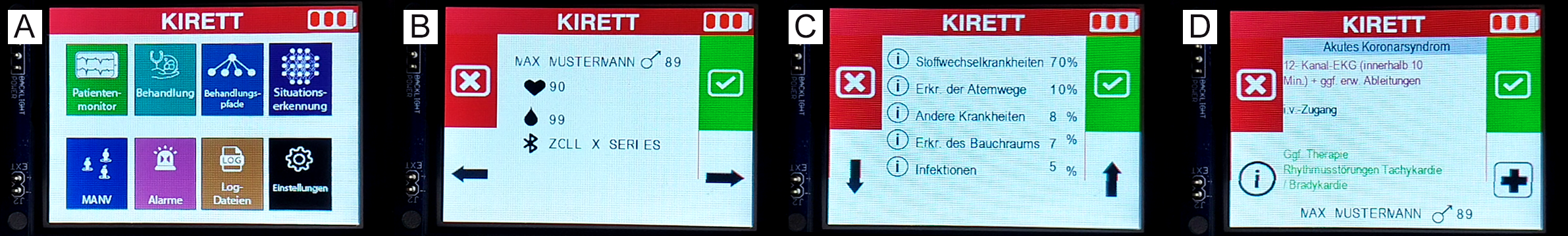}}
\caption{(A) main menu, (B) patient monitor, (C) situation detection and (D) treatment screen}
\label{screens}
\end{figure*}

The treatment screen is used as the central screen to allow knowledge graph navigation. Beforehand the initial treatment path selection by health professionals needs to be manually selected to provide the correct treatment path start node. In the background, situation detection and data retrieval from medical devices is triggered continuously. Figure \ref{screens} (D) shows four buttons, which enable navigation through the knowledge graph. The info button on the left button screen part enables to see further supporting information for the current task. The cross button on the bottom-right corner visualizes the patient monitor.

\begin{comment}
\begin{figure}[htbp]
\centerline{\includegraphics[width=0.40\textwidth]{images/treatment.png}}
\caption{Treatment Screen}
\label{treatmentscreen}
\end{figure}
\end{comment}

 The screen is divided into five sections. The first presents the selected treatment path, to remind the user which treatment path is selected. The further three sections are visualized in three different colors (red, black, green). They describe the previous, current, and next treatment steps. This supports the user to revisit their last decision or action and also look into the "future" to confirm medical actions. The colors red and green present the button colors to enable second visual support for the user experience.

\subsection{Patient Monitor}
Health professionals observe medical data throughout an emergency scenario to be able to react to changes in the vitals. The patient monitor (fig. \ref{screens} (B)) is used to visualize vitals, like pulse and oxygen saturation, along with the patient's name, sex, and age.

\subsection{Massive Information Visualization}
The knowledge graph also provides the possibility to visualize further and more detailed information with an intention of orientation and overview. For that, the smallest possible font size will be used. The screen only contains the text and one button to go back to the main treatment screen. 

This screen will provide a large area where text can be visualized freely to represent the graph. How text will be prepared for visualizing information fast and comprehensively, will be developed in the ongoing project.

\subsection{Element-selecting screen}
During an emergency situation, a situation detection algorithm is constantly reevaluating available information on the base of recorded data from medical devices. This situation detection results are represented as a vector of five possible situations (see section \ref{sd}. In addition, the knowledge-graph-based treatment inference also needs manual input to navigate the treatment graph. 

\begin{comment}
\begin{figure}[htbp]
\centerline{\includegraphics[width=0.4\textwidth]{images/situationdetection.png}}
\caption{Situation Detection Screen}
\label{situationdetection}
\end{figure}
\end{comment}

Figure \ref{screens} (C) presents an element selecting screen for a calculated situation detection. Health professionals can select from the given "illness groups" the most suitable explanation for the patient's current situation and select it via the green "acceptance"-button. \\

\subsection{Main Menu}
The main menu (fig. \ref{screens} (A)) is additionally implemented as a core piece of navigation through the different visual elements. It is used to visualize all different functionalities for direct access within a rescue operation. The elements are listed in a grid layout with different colors to enable an easier differentiation of the elements. The elements contain visuals, containing the patient monitor  (fig. \ref{screens} (B)) which shows relevant information about the patient, the ongoing treatment  (fig. \ref{screens} (D)), all existing treatments, the opportunity to manually trigger the situation detection  (fig. \ref{screens} (C)), the MCI scenario (Mass Casualty Incident), alarms, logging-screens, and the settings. The MCI scenario will be highlighted in future publications. 

\begin{comment}
\begin{figure}[htbp]
\centerline{\includegraphics[width=0.40\textwidth]{images/mainmenu.png}}
\caption{Main Menu Screen}
\label{mainmenu}
\end{figure}
\end{comment}

\section{Evaluation}

The requirements analyzed, implemented, and presented in this paper describe a UX-design for an emergency rescue wearable for health professionals. All screen components were evaluated by health professionals of associative partners from the German Red Cross Siegen (DRK) and the Jung-Stilling Hospital in Siegen in multiple sessions, checking if buttons, sizes, colors, and other elements are sufficient to use. In future, the wearable will be trialed in a simulated rescue operation scenario, where the reaction time, the user experience, and the hardware itself is tested. 
As a future experimental perspective, the test environment will be enhanced with an audio-navigation component, which enables end users to use voice commands to navigate through the knowledge graph component. 

\section{Conclusion}
This paper presents the result of the visual component development in the KIRETT project and provides a possible UX-design for rescue operations in Germany. The paper started by presenting the need for a UX-design in rescue operations and provided a list of requirements. 
Then the paper presents some essentials of the understanding of UX-design and the utilized knowledge graph and presents the LCD and FPGA-board used in the KIRETT project, along with limitations of the wearable implementation. Furthermore, this paper developed methods for communication across processors and a text-to-image generator to overcome the limitations of the used LCD. The paper showed how requirements retrieved from the knowledge graph and medical personnel of rescue operations can be combined to present a smart UX-design for rescue operation wearable which enables easy information retrieval during an emergency situation. 

\section*{Funding}
Financial support for this ongoing research is provided by the Federal Ministry of Education and Research in Germany. This research has been supported by the KIRETT project coordinator CRS Medical GmbH (Aßlar, Germany) and partner mbeder GmbH (Siegen, Germany). Furthermore, the authors give their gratitude to the associative partners: Kreis Siegen Wittgenstein, City of Siegen, the German Red Cross Siegen (DRK) and the Jung-Stilling-Hospital in Siegen.

%Bibliography
\bibliographystyle{unsrt}  
\bibliography{templateArXiv}

\end{document}